\documentstyle[11pt,paspconf]{article}

\begin{document}

\title{Simulations of Ly$\alpha$ Absorption from Low Surface Brightness
Galaxies}
\author{Suzanne M. Linder}
\affil{Pennsylvania State University, University Park, PA 16802 USA;
 slinder@astro.psu.edu}

\begin{abstract}
Using simulations of the low redshift galaxy population based upon galaxy
observations, it is shown (Linder 1998) that the majority of Ly$\alpha$ 
absorbers
at low redshift could arise in low surface brightness (LSB) galaxies.
The contribution to absorption from LSB galaxies is large for any 
galaxy surface brightness distribution which is currently supported by
observations.  Ly$\alpha$ absorbers should become powerful tools for 
studying the properties and evolution of galaxies, but first it will be
necessary to establish observationally the nature of the Ly$\alpha$ absorbers
at low redshift.  Further simulations, in which the absorbing galaxy 
population is 'observed' with some selection criteria, are used to explore
how easily possible it is for an observer to test for a scenario in which
LSB galaxies give rise to most of the Ly$\alpha$ absorbers.  It is shown
that absorption arising in LSB galaxies is often likely to be attributed
to high surface brightness galaxies at larger impact parameters from the
quasar line of sight.

\end{abstract}

\keywords{quasar absorption lines, galaxies}

\section{Introduction}

\begin{figure}[tb]
\plotfiddle{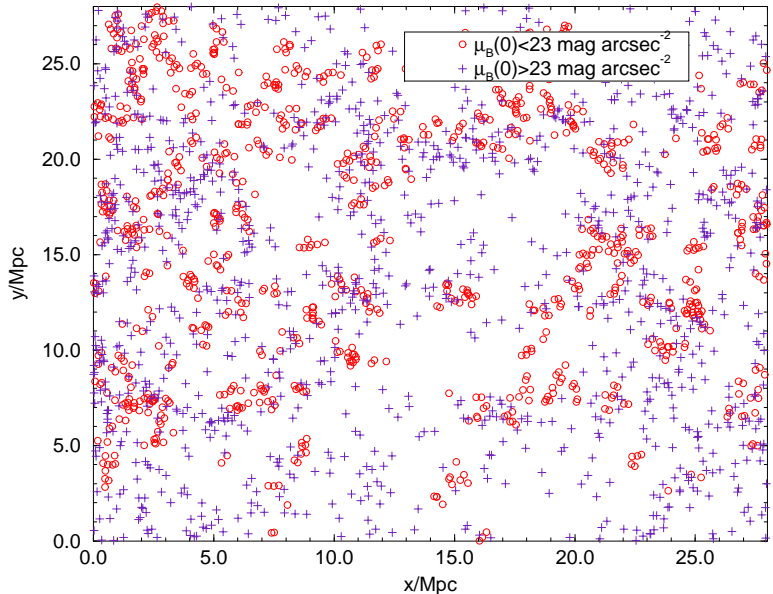}{3.2in}{270}{50}{50}{-185}{255}
\vskip  +0.1in
\caption{A slice of the cluster simulation which is 5 Mpc thick is 
shown, illustrating the positions of LSB and HSB galaxies.  LSB galaxies
are made to be clustered more weakly than HSB galaxies.}
\label{fig1}
\end{figure}

\begin{figure}[bt]
\plotfiddle{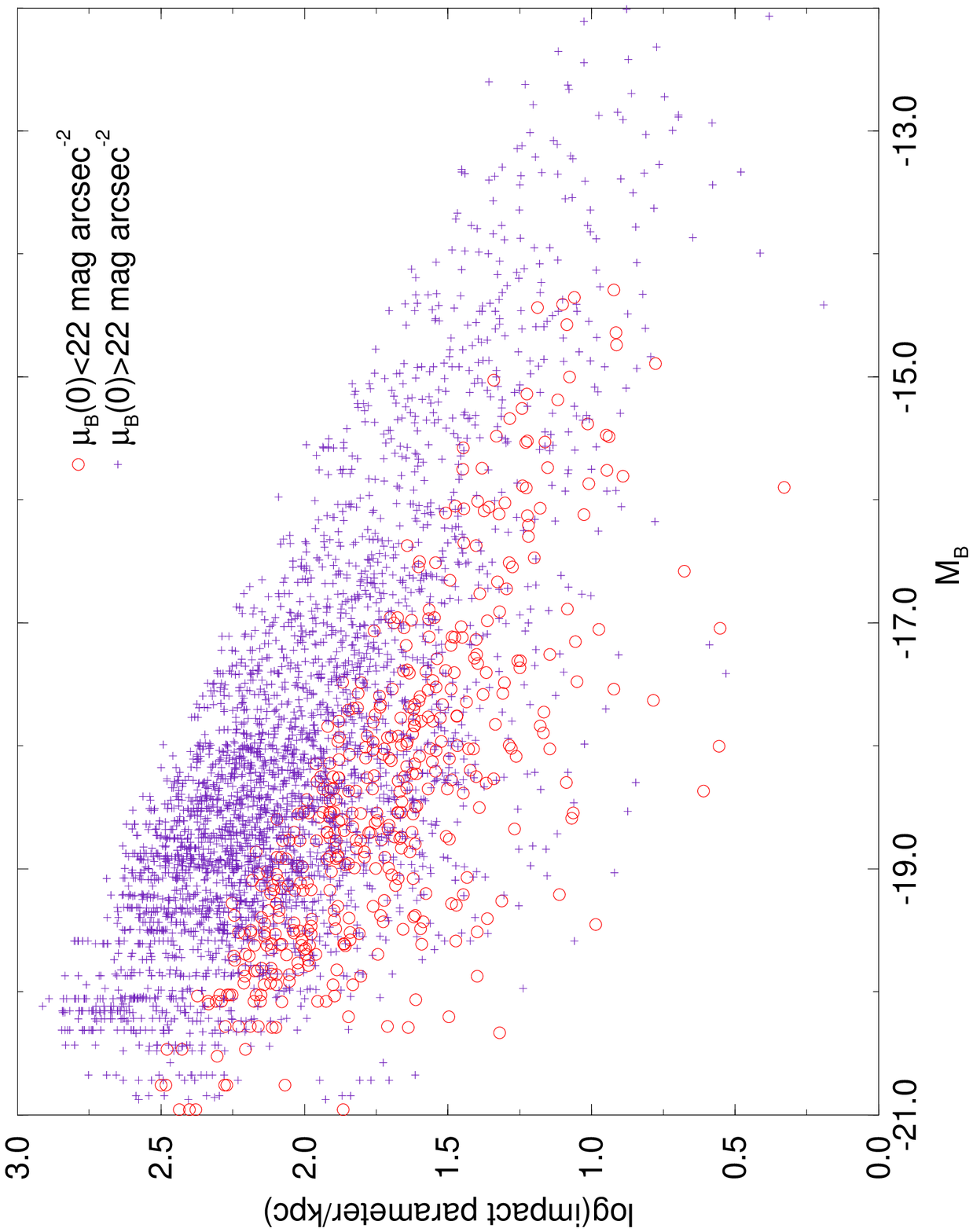}{3.2in}{270}{50}{50}{-185}{255}
\vskip +0.1in
\caption{The impact parameter (between the galaxy center and the quasar
line of sight) is plotted versus $M_B$ for the actual absorbing ($>
10^{14.3}$
cm$^{-2}$) galaxies
in the simulations.  More luminous galaxies have larger absorption radii,
and can thus cause absorption at larger impact parameters from the line of
sight.  LSB galaxies have larger absorption radii than HSB galaxies at a 
given luminosity.}
\label{fig2}
\end{figure}

Ly$\alpha$ absorption at low redshift allows for a powerful probe of gas in or 
around galaxies.  With Ly$\alpha$ absorber observations it is possible to
detect neutral hydrogen column densities which are several orders of 
magnitude lower than those typically seen with 21 cm observations.  For
example, the HST Key Absorption Line project (Bahcall et al. 1996) is 
complete to approximately $N_{HI}\sim 10^{14.3}$ cm$^{-2}$.  Observations of
Ly$\alpha$ absorption should therefore be capable of detecting neutral gas
which is known to exist in galaxy disks as well as more diffuse gas or 
neutral components of highly ionized gas surrounding galaxies.  Given that
Ly$\alpha$ absorption lines can be observed even more easily at higher
redshifts, they should become especially useful tools for studying galaxy
evolution in the future.

\begin{figure}[thb]
\plotfiddle{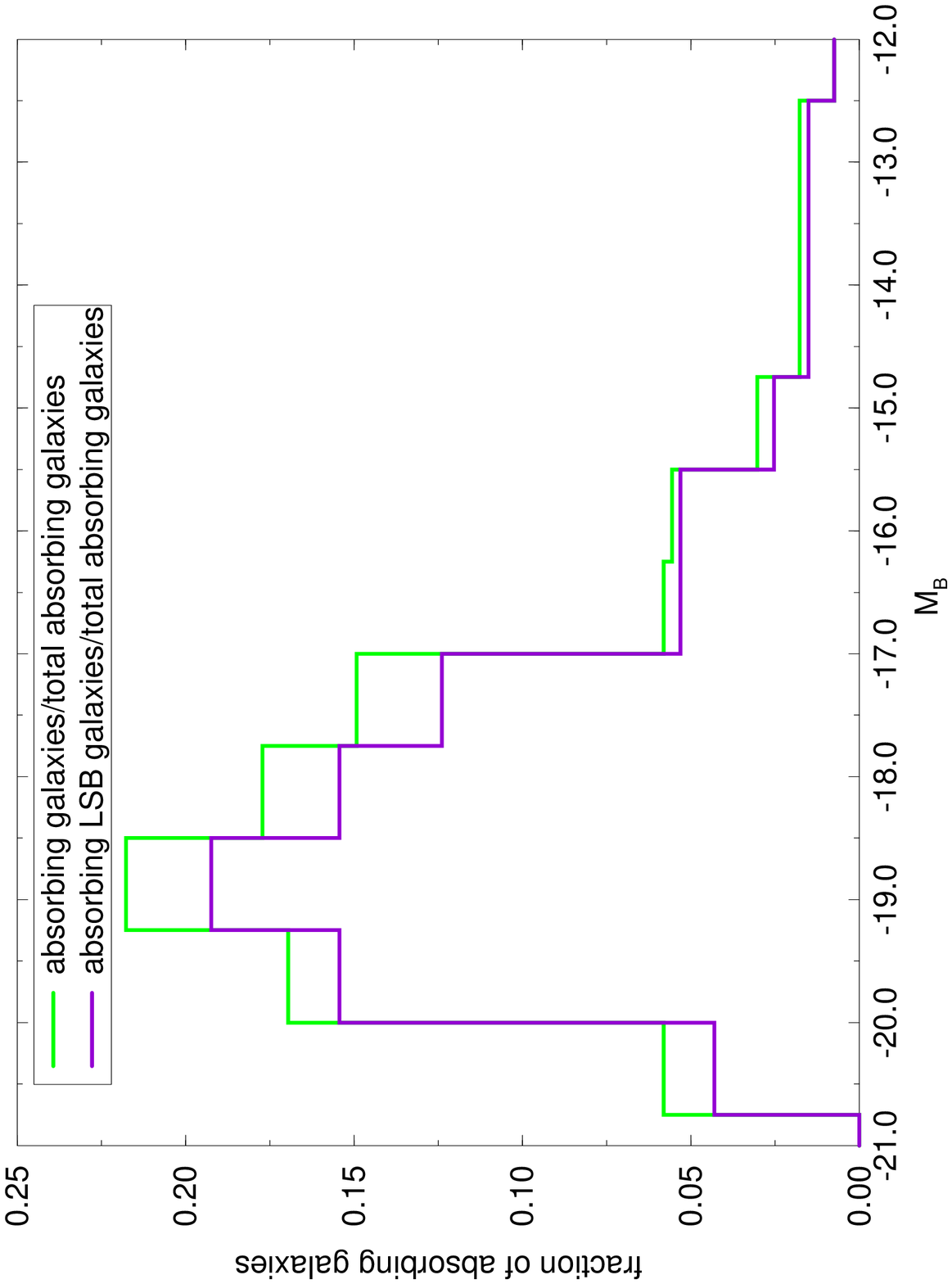}{3.2in}{270}{50}{50}{-185}{255}
\vskip +0.1in
\caption{A distribution of $M_B$ is shown for the actual absorbing ($>10^{15}$
cm$^{-2}$) LSB ($>22$ B mag arcsec$^{-2}$) galaxies and all galaxies, both
normalized to the total number of absorbing galaxies.  Moderately luminous
LSB galaxies are shown to cause most of the Ly$\alpha$ absorption.}
\label{fig3}
\end{figure}

\begin{figure}[tb]
\plotfiddle{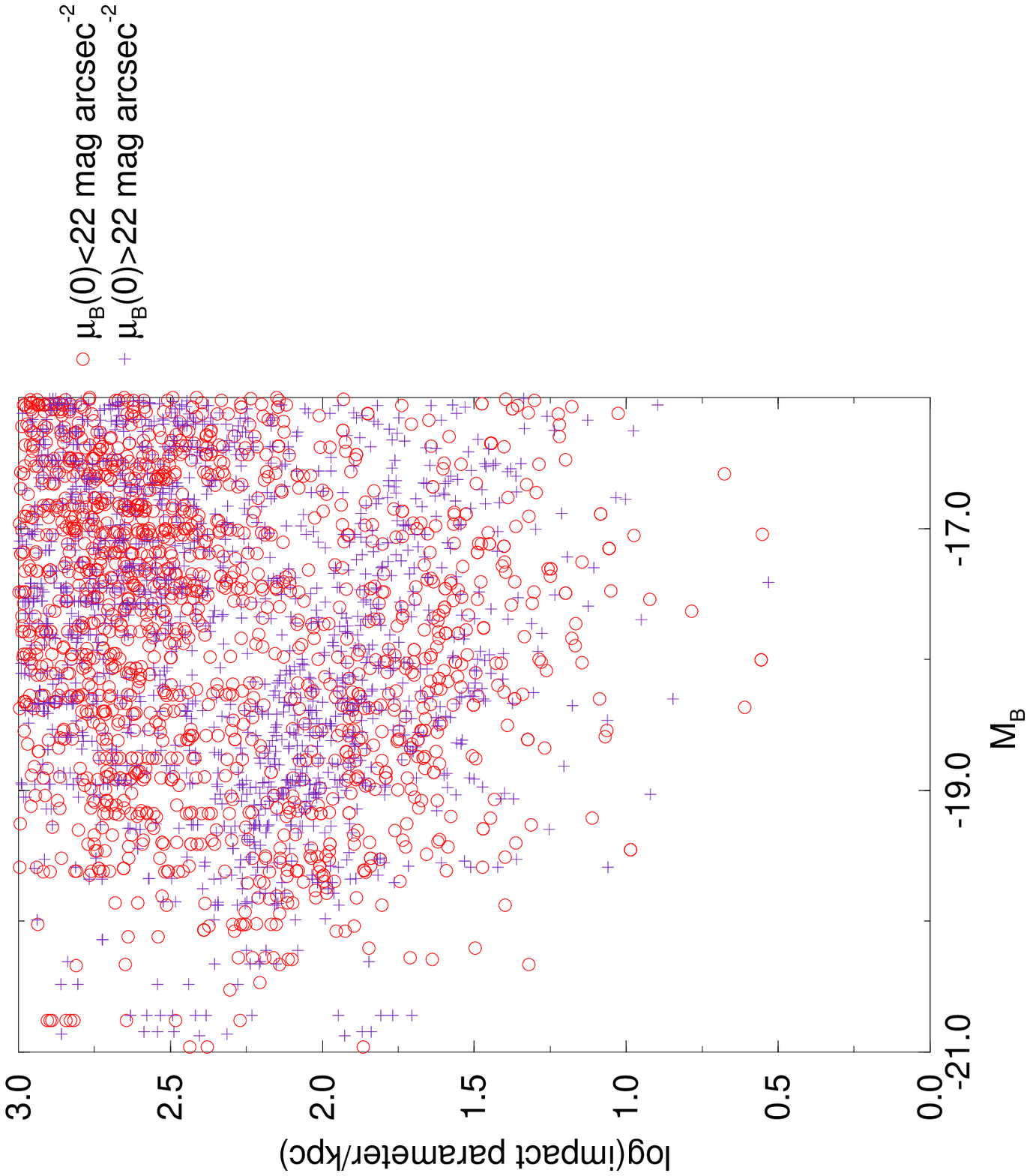}{3.2in}{270}{50}{50}{-185}{255}
\vskip +0.1in
\caption{The impact parameter versus $M_B$ are shown as if they were obtained
by 'observing' the simulated galaxies, in order to identify an absorbing 
galaxy, according to the selection criteria given in the text. While the
same simulated galaxies shown in Fig. 2 are 'observed' here, it
is no longer obvious that LSB galaxies have larger absorption cross sections.}
\label{fig4}
\end{figure}

Some stronger Ly$\alpha$ absorption lines are known to arise in lines of 
sight thorough galaxies, but the nature of the more common, weak forest
absorbers is more difficult to establish.  Supposing that these absorbers
arise in gas associated with galaxies,  the galaxies may be located at 
large impact parameters from a quasar line of sight.  Therefore it becomes
impossible to be sure that any particular galaxy is actually causing the
absorption, so that it is difficult to make a direct test of any model
in which galaxies cause absorption.

Testing absorber-galaxy models becomes even more complicated if absorption
arises in unidentified galaxies.  Given that low surface brightness (LSB)
galaxies are common relative to high surface brightness (HSB) galaxies,
and that LSB galaxies are often found to be rich in gas, they {\it must} make 
an important contribution to Ly$\alpha$ absorption.  LSB galaxies are 
typically bigger than HSB galaxies at a given luminosity.  If LSB galaxies
also have larger sizes compared to HSB galaxies in Ly$\alpha$ absorption, then
they may easily dominate the cross section for Ly$\alpha$ absorption.
Ly$\alpha$ absorbers may allow for an important way to study the properties
and evolution of galaxies, especially LSB galaxies, but first it will
be necessary to find a way to test absorber-galaxy models and to establish 
what types of galaxies could actually cause most of the Ly$\alpha$ absorption.

\begin{figure}[tb]
\plotfiddle{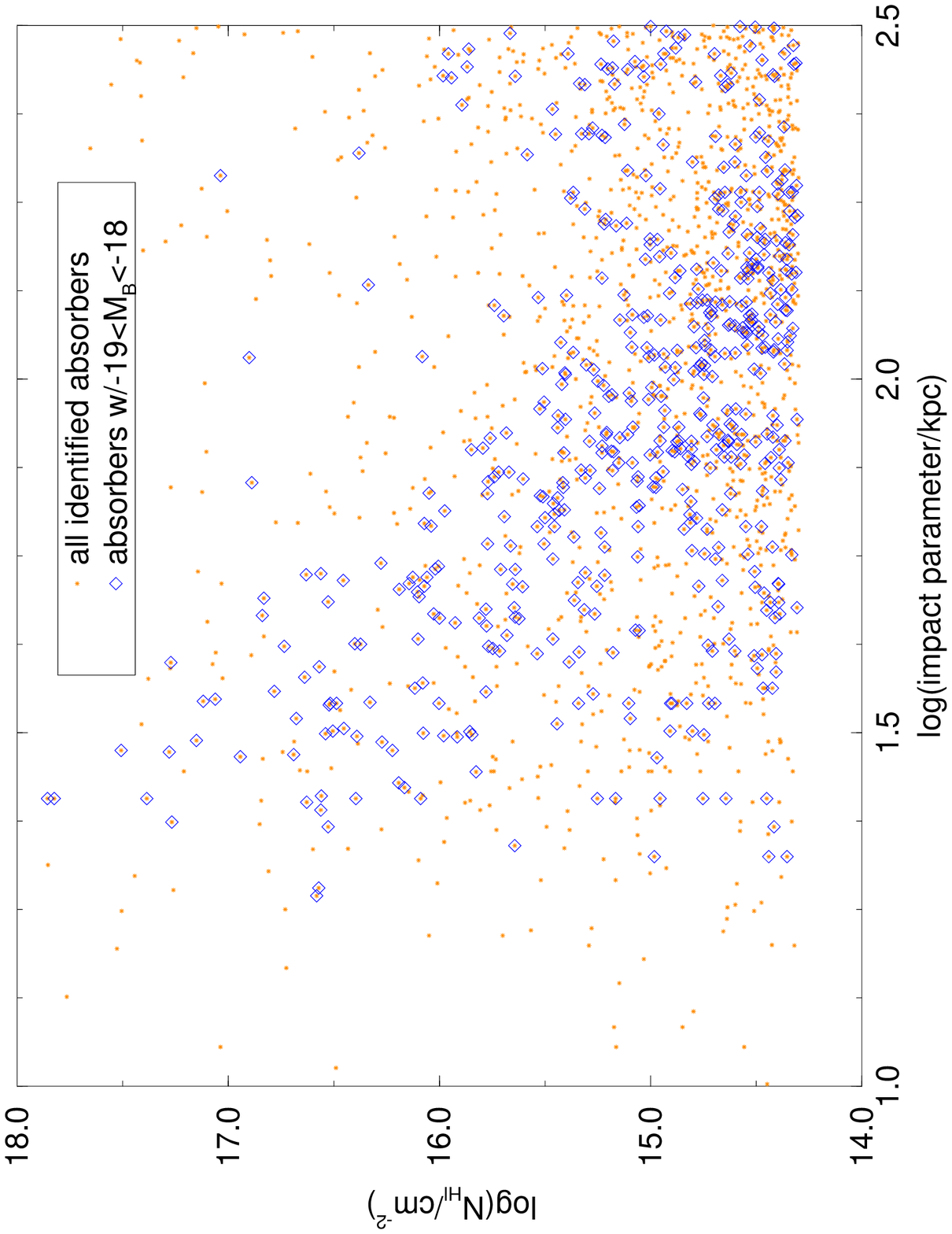}{3.2in}{270}{50}{50}{-185}{255}
\vskip +0.1in
\caption{The neutral column density for Ly$\alpha$ absorption is plotted 
versus the impact parameter between galaxy and line of sight for the
'observed' (see text) absorbing galaxy.  An anticorrelation between 
neutral column density and impact parameter is still seen even though
the actual absorbing galaxy is unlikely to be identified (particularly
for the weaker absorption lines). }
\label{fig5}
\end{figure}

\begin{figure}[tb]
\plotfiddle{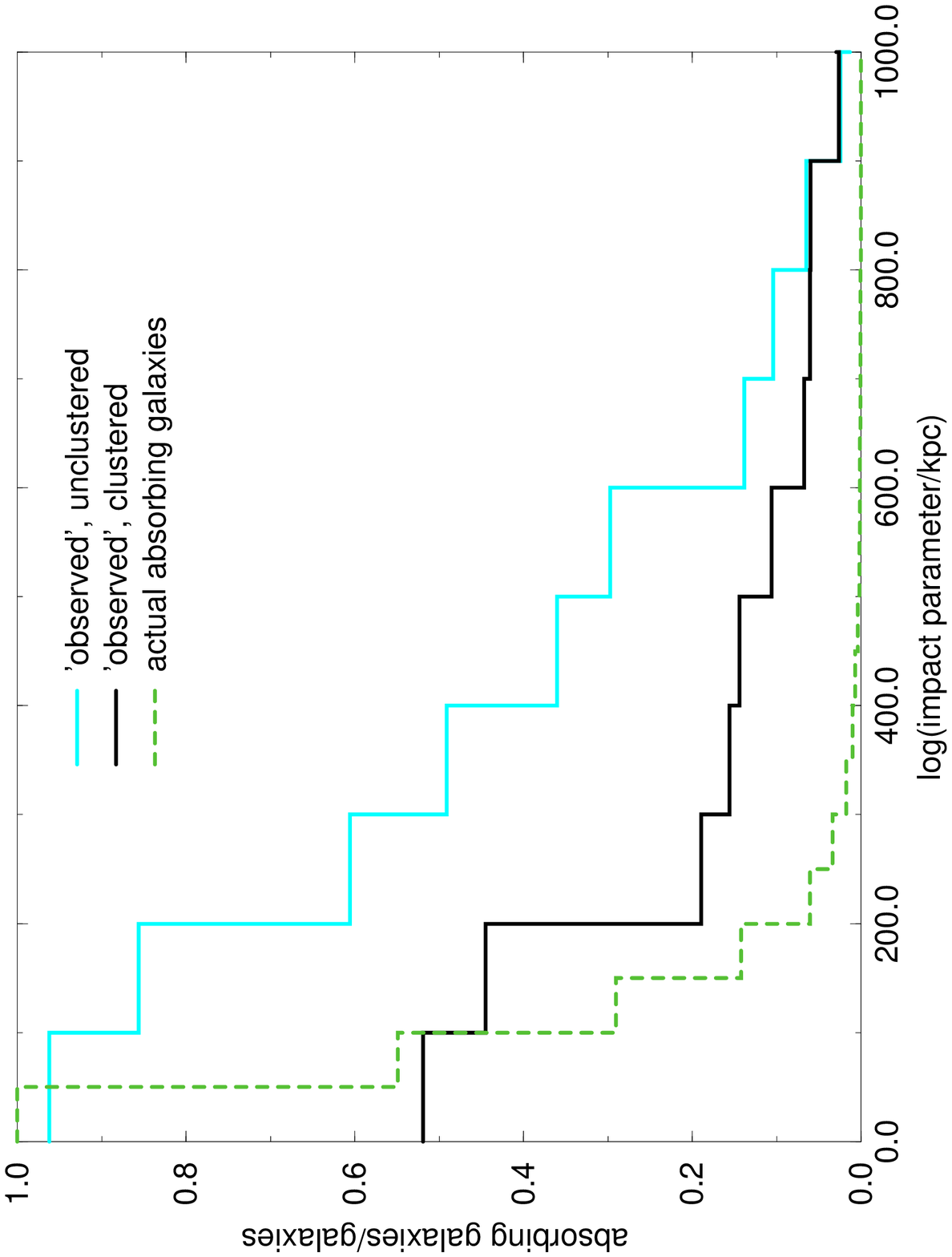}{3.2in}{270}{50}{50}{-185}{255}
\vskip +0.1in
\caption{A 'covering factor' plot shows the fraction of galaxies found 
at some impact parameter from a line of sight which cause absorption
($>10^{14.3}$ cm$^{-2}$).   Galaxy absorption radii are likely to be overestimated
when the galaxies are observed with strong selection effects against LSB 
galaxies.  Galaxy clustering also causes misidentification of the actual
absorbing galaxy to occur more frequently.}
\label{fig6}
\end{figure}

\section{Simulations of Absorbing Galaxies}

Simulations are used (Linder 1998) to find out how many absorbers 
could be associated with galaxies.  A population of galaxies is defined
according to observed distributions of galaxy parameters, and the galaxies
are placed in a box.  Random lines of sight are chosen to go through the 
box, and absorption occurs where the lines of sight 
intersect the galaxies.
In these simulations it is assumed that absorption arises in galaxy disks
and their ionized outer extensions (Charlton, Salpeter \& Linder 1994).
It is shown that Ly$\alpha$ absorber counts at low redshift
can easily be explained by galaxies when LSB galaxies are included.
The model parameters are tuned to be consistent with 
observed luminosity functions for HSB and 
LSB galaxies, standard nucleosynthesis predictions of the baryon density, and
Lyman limit absorber counts.  Preferred scenarios are found to have moderate
galaxy absorption cross sections and moderate numbers of LSB galaxies.

The galaxies in the earlier simulations are placed
randomly in the box.  More recently, clustered galaxies have been simulated
using a fractal-type method as in Soniera \& Peebles (1978) where eight
levels of clustering are used.  The LSB galaxies are moved further out
from the centers of the clusters at the second largest level, so that they
are made to be clustered more weakly than HSB galaxies, as shown in a
slice of the box in Fig.~\ref{fig1}.

More luminous galaxies have larger absorption radii in the simulations, 
so that they are able to cause absorption when they are at  
larger impact parameters from quasar lines of sight, as shown in
Fig.~\ref{fig2}.
The Holmberg relation ($R\sim L^{0.5}$) and the typical absorbing radius
of a luminous HSB galaxy are similar to those reported in some observational
studies (Chen et al. 1998; Lanzetta, this {\it Proceedings}).
It can also be seen from Fig.~\ref{fig2} that LSB galaxies have larger absorption 
radii than HSB galaxies at a given galaxy luminosity.  Thus the majority
of Ly$\alpha$ absorbers arise in lines of sight through LSB galaxies, as
seen in Fig.~\ref{fig3}.  It can also be seen in Fig.~\ref{fig3}
that the Ly$\alpha$ 
absorption is generally caused by moderately luminous, 'normal' LSB galaxies
and not by extremely luminous Malin-type objects.  

A flat central surface brightness distribution (McGaugh 1996) at a given 
scale length was assumed for most of the simulations.  Preliminary
simulations have also been done using a surface brightness distribution 
which is lognormal at a given luminosity (de Jong, this {\it Proceedings}).  
While this surface brightness distribution allows for fewer extremely
luminous LSB galaxies, there are still numerous moderately luminous LSB
galaxies which still give rise to most of the Ly$\alpha$ absorption.

\section{'Observing' the Simulations}

Given some scenario where LSB galaxies make an significant contribution 
to Ly$\alpha$ absorption, such as the one described above, is it possible
for an observer to test for such a scenario?  Suppose we can 'observe' the
galaxies simulated above using some selection criteria.  Here the nearest
galaxy to a line of sight is found within a velocity difference of 400 km 
s$^{-1}$ to the absorption line, where the galaxy has $M_B<-16$ and $\mu_
B(0)<23$ mag arcsec$^{-2}$.

A plot of $M_B$ versus the impact parameter of the 'observed' absorbing 
galaxy is shown in Fig.~\ref{fig4}, using the same scale as the actual plot shown
in Fig.~\ref{fig2}.  It is likely that an observer could find a way of determining
that the points in the upper right-hand corner for Fig.~\ref{fig4} are unphysical
absorber-galaxy associations.
For the remaining points, a correlation 
between luminosity and impact parameter can still be seen, although the
slope of the 'observed' Holmberg relation may change.
However, absorption arising in LSB galaxies is frequently 'observed' as 
arising in HSB galaxies at typically larger impact parameters from the 
quasar line of sight.  Thus it is no longer possible to verify from the 
'observed' plot that LSB galaxies have larger absorption cross sections, 
as assumed in the simulation.

In any reasonable absorber-galaxy model, it is expected that, on average,
the neutral hydrogen column density should fall off with galaxy radius.
Thus it is likely that absorbing galaxies at larger impact parameters 
from the quasar line of sight should produce lower column densities.  An
anticorrelation between impact parameter and neutral column density
is seen for the simulated actual absorbing galaxies (Linder 1998).  
However, a large amount of scatter exists in such plots, which is caused
mostly by variations in the properties of absorbing galaxies.  (Varying
disk inclinations produce comparatively little scatter, so that a 
spherical halo-type model for absorbing galaxies should produce a similar
plot.) 

The anticorrelation between neutral column density and impact 
parameter can still be seen for the galaxies 'observed' with the selection
criteria above, as shown in Fig.~\ref{fig5}.  Note that if more reasonable 
selection effects were used to 'observe' the simulations, compact HSB
dwarf galaxies would be excluded, so that fewer points would be seen 
toward the lower left-hand corner.   Thus, when also considering
incompleteness for observing galaxies at large impact parameters, 
Fig.~\ref{fig5}
should 
bear even more resemblance to the plot shown in Chen et al. (1998).
LSB absorbers are 
frequently misidentified in the simulation shown.  In fact, many LSB 
absorbers have no possible 'observed' galaxy which satisfies the selection
criteria, especially when the LSB galaxies are clustered more weakly than
the HSB galaxies.  Thus an observer would be likely to conclude that some 
fraction of absorbers cannot be explained by galaxies.  Less scatter is
seen in Fig.~\ref{fig5} for galaxies with a narrow range of luminosities.  However,
it is more difficult to reduce the scatter by limiting the range of
'observed' absorbing galaxy surface brightnesses.  Thus from this plot 
an observer is likely to conclude again that galaxy absorption cross
sections are correlated with galaxy luminosity but not with galaxy surface
brightness. 

Another plot frequently made by observers (for example, Bowen, Blades, \& 
Pettini 1996), that illustrating 
the absorption covering factor, or the fraction of galaxies found to 
cause absorption as a function of impact parameter, is shown in Fig.~\ref{fig6}.
'Observed' galaxies appear to cause absorption at large impact parameters
compared to those seen in Fig.~\ref{fig2}.  Again it can be seen that many absorbers
arising in LSB galaxies are attributed to HSB galaxies at larger impact
parameters from the quasar line of sight.   It can also be seen that 
clustered galaxies are more frequently misidentified.  It will be difficult
to use an observed covering factor plot to test an absorber-galaxy model
since its appearance will be sensitive to the actual absorbing properties
of the galaxies, the clustering behavior of the galaxies, and the 
observational selection criteria.

\section{Conclusions}

It has been shown that Ly$\alpha$ absorbers at low redshift
can easily be explained by galaxies when LSB galaxies are included.  The
majority of absorbers are likely to arise in LSB galaxies, even if 
extremely luminous LSB galaxies are rare.
Ly$\alpha$ absorber observations will be useful for constraining properties
of galaxies, such as the gaseous extent of galaxies, and for studying 
evolution in such galaxy properties. Ly$\alpha$ absorbers will be especially 
powerful tools for studying galaxies which are LSB at any evolutionary stage.

In order to take advantage of the information from Ly$\alpha$ absorption,
it will be necessary to establish observationally specifically what kinds
of galaxies give rise to Ly$\alpha$ absorption.  Observational studies 
which attempt to match absorption lines with possible absorbing galaxies
will be crucial to establishing the nature of the Ly$\alpha$ absorbers.  
However, as shown above, it is also necessary to use simulations in order
to understand the absorbing properties of galaxies.  For example, an 
observer may overestimate the absorbing radius of a galaxy by looking 
directly at observations which are subject to severe selection effects.
Selection effects will always be present in observational studies, but
it should be possible to define the selection criteria
more rigorously so that they can be simulated in order to make 
more realistic tests of absorber-galaxy models.

\acknowledgments  I am grateful to J. Charlton and C. Churchill for helpful
discussions, R. de Jong for supplying a surface brightness distribution
before publication, WISE CIC and NASA GSRP for travel funding, and
R. Gunesch and A. Panaitescu for assistance with figures.

\end{document}